# Gravity-induced phase phenomena in plate-rod colloidal mixtures

Tobias Eckert [1], Matthias Schmidt [1]✉ & Daniel de las Heras [1]✉

Gravity can affect colloidal suspensions since for micrometer-sized particles gravitational and thermal energies can be comparable over vertical length scales of a few millimeters. In mixtures, each species possesses a different buoyant mass, which can make experimental results counter-intuitive and difficult to interpret. Here, we revisit from a theoretical perspective iconic sedimentation-diffusion-equilibrium experiments on colloidal plate-rod mixtures by van der Kooij and Lekkerkerker. We reproduce their findings, including the observation of five different mesophases in a single cuvette. Using sedimentation path theory, we incorporate gravity into a microscopic theory for the bulk of a plate-rod mixture. We also show how to disentangle the effects of gravity from sedimentation experiments to obtain the bulk behavior and make predictions that can be experimentally tested. These include changes in the sequence by altering the sample height. We demonstrate that both buoyant mass ratio and sample height form control parameters to study bulk phase behavior.

[1] Theoretische Physik II, Physikalisches Institut, Universität Bayreuth, Bayreuth, Germany. ✉email: matthias.schmidt@uni-bayreuth.de; delasheras.daniel@gmail.com

Colloids, i.e., nano- to micrometer-sized particles suspended in a liquid, behave under some circumstances as big atoms[1–5]. Despite their size disparity, molecular and colloidal systems exhibit analogous bulk phases and similar surface phenomena such as wetting[6], capillary waves at the free fluid-fluid interface[7], and the occurrence of topological defects due to frustration[8]. Beyond intrinsic fundamental interest, understanding bulk phase behavior is a prerequisite for the design of new materials. Sedimentation experiments, in which a colloidal suspension is placed in a cuvette under the influence of gravity, are ideal candidates to study the bulk behavior of colloidal systems. One famous example are the experiments on suspensions of nearly hard colloidal spheres[9] that confirmed the fluid-crystal transition predicted decades earlier[10]. Sedimentation experiments also confirmed the entropy-driven formation of liquid crystalline phases in systems of anisotropic particles[11,12], the existence of empty liquids and equilibrium gels[13], and the hexatic phase in two-dimensional colloidal discs[14]. Gravity can play a major role in the colloidal realm and can only be neglected if the relevant length scales of the experiment (e.g., the height of the sample inside the cuvette) are much smaller than the colloidal gravitational length; the latter is the ratio between the thermal energy and the buoyant force acting on a single particle. For typical colloidal systems, the gravitational length is of the order of millimeters, and hence it is smaller or comparable to the sample height.

Depending on the sign of the buoyant mass the colloidal particles sediment towards the bottom or they cream up. In both cases, gravity creates a particle density gradient in the vertical direction. Sedimentation-diffusion-equilibrium is reached once the gravity-induced particle flow is balanced by the diffusive flow originated by the density gradient and the interparticle interactions. As already shown in Perrin's pioneering experiments[15], the resulting height-dependent colloidal density distribution provides direct access to the full equation of state for monocomponent systems of both isotropic[16,17] and anisotropic[18,19] colloidal particles.

It is often an excellent approximation to consider that at each height inside the cuvette, the system is well reproduced by a corresponding homogeneous equilibrium system with a bulk density that is identical to the local density of the inhomogeneous system[20,21]. This local density approximation can be implemented by considering that the chemical potential of the sample varies linearly with the vertical coordinate. The strength of sedimentation-diffusion-equilibrium experiments is that instead of looking at a single state point of a bulk system (i.e., at a fixed chemical potential), one is able to consider set of states with varying chemical potential along the vertical axis. This result is due to the gravity-induced varying density (or equivalently chemical potential) along the vertical axis. Since in many colloidal systems the gravitational length is smaller than the typical height of a cuvette, even sedimentation experiments with samples of few millimeters in height provide in-depth insight into the equation of state and bulk phase phenomenology.

In binary colloidal mixtures, gravity has stronger impact since two, in general distinct, gravitational lengths exist. Counterintuitive and complex phenomenology arises, making it difficult to draw conclusions about bulk behavior. For example, in their iconic experiments on plate-rod mixtures, van der Kooij and Lekkerkerker[22,23], found hitherto unexpected and rich phenomenology of colloidal mixtures. By changing the colloidal concentrations the authors observed the formation of different stacking sequences, including samples with the sequence: isotropic-nematic-smectic-nematic-columnar, when scanned from top to bottom of the sample. The two nematic layers correspond to different bulk phases rich in either rods (top) or plates (bottom).

In mixtures, it is frequent to observe more than three layers of different bulk phases in a cuvette at different altitudes: Up to six different layers occur in mixtures of positively charged colloidal plates and nonadsorbing polymers[24]. Even the same layer can reenter the stacking sequence, such as e.g., a nematic sandwiched between two isotropic layers in sphere-plate colloidal mixtures[25]. Further experimental studies were aimed at colloidal rod-plate[26,27], plate-sphere[28,29], rod-sphere[30], and sphere-sphere[31,32] mixtures, liquid crystalline binary nanosheet colloids[33], mixtures of thin and thick colloidal rods[34], as well as attractive nanosized spheres and plates[35]. Gravitational effects can be relevant even if the system contains only a few colloidal layers[36], as e.g in the stratification found in drying films of colloidal mixtures[37].

To draw conclusions about bulk phenomena from sedimentation experiments (and vice versa), gravity needs to be considered. However, in mixtures the dimensionality of the phase diagram increases by one unit for each added species. Gravity induces a height-dependent density profile for each component. Therefore the gravity-induced one-dimensional scan along the vertical axis of the sample gives only a one-dimensional slice of the complete phase diagram. The full equation of state and the phase diagram can not be extracted from a single sedimentation profile. Wensink and Lekkerkerker[20] incorporated gravity in a mixture of plates and polymers by treating the mixture as an effective mono-component system with the chemical potential of the polymer fixed. This approach is limited to systems in which the gravitational length of one species (the polymer) is much larger than the sample height. A generalized Archimedes principle[38,39] appropriately describes the behavior a mixture in which both species are colloidal particles and one of them is very diluted.

An alternative approach, valid for any mixture and any colloidal concentration, was formally given by de las Heras and Schmidt[21]. The theory is formulated in terms of sedimentation paths, which represent how the chemical potentials of both species vary linearly with the vertical coordinate due to gravity. The sedimentation paths are straight lines in the plane of chemical potentials, and the crossing between a path and a binodal indicates the formation of an interface in the sample. Different stacking sequences appear depending on which binodals are crossed by the path. The stacking sequences are grouped in a stacking diagram, which is the analog of the bulk phase diagram for systems subject to gravity. So far, the stacking diagrams have been used to theoretically study sedimentation of model colloidal mixtures[21,40–42].

We demonstrate here that this formal approach also opens the door for the rigorous interpretation and the prediction of sedimentation-diffusion-equilibrium experiments in colloidal mixtures. We reinterpret the findings of the arguably best known experimental study in the field, conducted by van der Kooij and Lekkerkerker[22,23], on plate-rod colloidal mixtures. By incorporating gravity into a microscopic theory for the bulk behavior of the mixture we reproduce quantitatively their experimental findings. Furthermore we address the (experimentally relevant) inverse problem. That is, we demonstrate how to infer the bulk phase diagram using the experimentally obtained stacking sequences and the individual heights of their constituent layers. We also make predictions that can be tested experimentally: a different set of stacking sequences emerges by altering the ratio of the buoyant masses of the colloidal particles and complex changes in the stacking sequence occur by simply varying the height of the sample. Both variables, the buoyant mass ratio and the sample height, can be systematically controlled in both experimental and theoretical work. Our demonstration of the important role played by both the buoyant mass ratio and the sample height allows to design experimental and theoretical studies that exploit these (hitherto largely unexplored) control parameters.

## Results

**Particle model.** To model the experiments[22,23] we consider a mixture of hard rods and hard plates. Depending on composition and packing fraction, the bulk phase is either isotropic (I), nematic rich in plates ($N_p$), or nematic rich in rods ($N_r$), see Fig. 1. We use subscripts p and r to designate the plates and the rods, respectively.

The gravitational lengths are $\xi_i = k_B T/(m_i g)$ with $m_i$ the buoyant mass of species $i$, $g$ the gravitational acceleration, $k_B$ the Boltzmann constant, and $T$ absolute temperature. The rods are made of boehmite (mass density 3.03 g/cm³) and the plates of gibbsite (2.35 g/cm³). The particles are sterically stabilized with a polymer coating of a few nanometer thickness and suspended in toluene (0.87 g/cm³). We use cylinders of lengths 200 nm and 10 nm, and diameters 20 nm and 150 nm to model the rods and the plates, respectively. The particles match both the length-to-width aspect ratio and the dimensions of the particles used in the experiments[22,23] within the experimental uncertainty. We subtract the volume of the polymer coating from the total particle volume to estimate the buoyant masses. Using a diameter of 15 nm for the rod's core and a length of 8.7 nm for the plates' core, we obtain $\xi_r = 5.5$ mm, $\xi_p = 1.8$ mm. Hence the buoyant mass ratio is

$$s = \frac{m_p}{m_r} = \frac{\xi_r}{\xi_p} \approx 3. \quad (1)$$

The values of the gravitational lengths and hence that of the buoyant mass ratio are only rough estimates since there is a large uncertainty in the particle dimensions (up to 25%) and in the thickness of the polymer layer[43,44]. Given this uncertainty, the buoyant mass ratio is likely between a minimum value of ~2 and a maximum value of ~5.

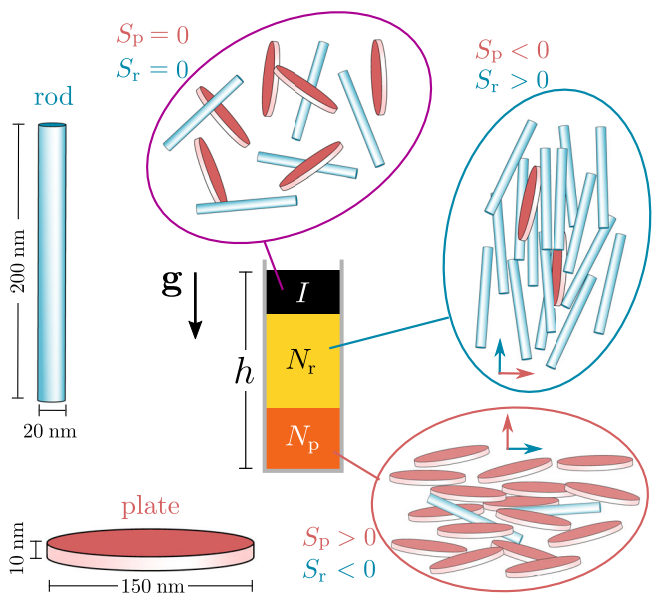

**Fig. 1 Model.** Schematics of colloidal rods (blue) and plates (red), together with schematics of the particles in the isotropic (I), the nematic rod-rich ($N_r$) and the nematic plate-rich ($N_p$) layers that appear in a cuvette of height $h$ under the gravitational field **g**. The blue and red arrows indicate the nematic director for rods and plates, respectively. The sign of the uniaxial order parameter of plates $S_p$ and rods $S_r$ in each phase is also indicated (the order parameter is calculated with respect to the director of the dominant species in each phase).

**Bulk.** We use an Onsager-like density functional theory to study the bulk, see Methods. To characterize the phases we use the uniaxial order parameters $S_i$ of each species (see Methods) that take values between −0.5 and 1. The order parameters measure the orientational order with respect to the direction given by the director of the dominant species, see Fig. 1. A positive (negative) value indicates that the alignment of the particles is parallel (perpendicular) to the director. In the isotropic phase both order parameters vanish.

**Sedimentation path theory.** Gravity is incorporated by approximating each horizontal slice of the system at height $z$ by a bulk equilibrium system with local chemical potentials $\mu_i(z)$ given by[21]

$$\mu_i(z) = \bar{\mu}_i - m_i g\left(z - \frac{h}{2}\right), \quad i = \text{r, p}, \quad (2)$$

with $0 \leq z \leq h$ the vertical coordinate and $h$ the height of the sample. This constitutes a local density approximation (LDA), which is justified if all correlation lengths are small compared to both gravitational lengths. The LDA is used only to incorporate gravity and it does not affect therefore the description of the bulk. Sophisticated bulk theories, such as fundamental measure theory[45], can be used together with sedimentation path theory to study sedimentation of e.g., crystalline phases.

Due to the gravitational potential, the local chemical potentials depend linearly on the vertical coordinate $z$. Equation (2) formalizes the concept that a sample subject to gravity can be understood as a set of bulk states at different chemical potentials and distributed along the vertical axis. Geometrically, Eq. (2) describes a line segment parametrized by $z$ in the plane of chemical potentials. We refer to such lines as sedimentation paths[21,41]. The constant terms $m_i gh/2$ in Eq. (2) conveniently translate the origin of chemical potentials such that the values of the midpoint of the path are $(\bar{\mu}_r, \bar{\mu}_p)$. Eliminating $z$ for the mixture in Eq. (2) yields

$$\mu_p(\mu_r) = s\mu_r + a, \quad (3)$$

which constitutes the equation of a line segment with a slope given by the buoyant mass ratio $s = m_p/m_r = \xi_r/\xi_p$ and intercept $a = \bar{\mu}_p - s\bar{\mu}_r$. The buoyant masses play therefore a vital role since they determine both the slope $s$ (buoyant mass ratio) and, together with $h$, the length of the sedimentation path in the plane of chemical potentials. The latter is given by $\beta\Delta\mu_i = h/\xi_i$, with $\Delta\mu_i = \mu_i(0) - \mu_i(h)$ being the differences in local chemical potentials between the bottom and the top of the sample.

The sedimentation path provides direct information of the sequence of layers in the sample, i.e., the stacking sequence. An interface between two layers of different bulk phases appears in the sample whenever a sedimentation path crosses a bulk binodal in the plane of chemical potentials, see Fig. 2a. The crossing point between the path and the binodal gives the $z$ position of the interface in the sample via Eq. (2).

**Stacking diagram.** Different stacking sequences occur by varying e.g., the position, the slope, and the length of the sedimentation path. The sequences can be grouped in a stacking diagram. The stacking diagram admits several representations depending on which variables are kept constant. To connect with the experiments we fix the buoyant masses and the sample height $h$. Hence, we work at constant sedimentation path length and fixed buoyant mass ratio $s$.

From the bulk phase diagram we construct the stacking diagram by finding the boundaries between two stacking sequences in the stacking diagram. There exist three types of boundaries formed by three sets of special sedimentation paths[41], see Fig. 2a. The first type corresponds to sedimentation paths that either start or end at a bulk binodal (red and yellow paths in Fig. 2). These paths represent boundaries, so-called sedimentation binodals of type I, between stacking sequences since an infinitesimal displacement of the path in the $\mu_1 - \mu_2$ plane can alter the stacking sequence by crossing the binodal. Paths that cross an ending point of a binodal, e.g., a triple point or a critical point (orange path in Fig. 2), also form a boundary in the stacking diagram, known as a terminal line. Finally, paths tangent to a binodal (blue path in Fig. 2) form the so-called sedimentation binodal of type II which is also a boundary between different stacking sequences. In each of the three cases, an infinitesimal displacement of the path can alter the stacking sequence.

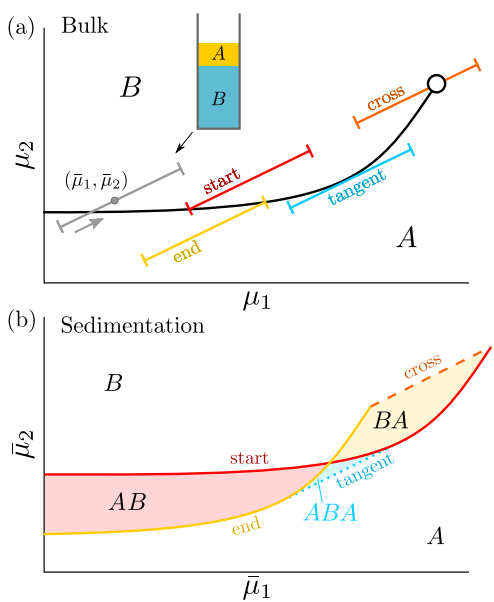

**Fig. 2 Sedimentation path theory. a** Model bulk phase diagram in the plane of chemical potentials $\mu_1 - \mu_2$. Phases $A$ and $B$ coexist along a binodal (solid line) that ends at a critical point (empty circle). The line segments are sedimentation paths. The gray path that crosses the binodal corresponds to a stacking sequence $AB$ (from top to bottom), as schematically represented. The gray arrow indicates the direction of the paths from top to bottom of the sample. Examples of all types of paths that form the boundaries between different stacking sequences are shown: (i) paths that either start (red) or end (yellow) at the binodal, (ii) paths tangent (blue) to the binodal, and (iii) paths crossing (orange) the critical point. An infinitesimal displacement of any of such paths can alter the stacking sequence. **b** Corresponding stacking diagram in the plane of average chemical potentials $\bar{\mu}_1 - \bar{\mu}_2$. Each colored region is a different stacking sequence, as indicated. The boundary lines between sequences are sedimentation binodals of type I (solid lines) or type II (dotted-blue line), and terminal lines (dashed-orange line).

The coordinates of the midpoint $(\bar{\mu}_r, \bar{\mu}_p)$ for each special path are then represented in the $\bar{\mu}_r - \bar{\mu}_p$ plane to generate the stacking diagram, see Fig. 2b. The experimentally relevant variables are the colloidal packing fractions. Therefore, we transform the stacking diagram from the $\bar{\mu}_1 - \bar{\mu}_2$ to the $\bar{\eta}_1 - \bar{\eta}_2$ plane of average colloidal packing fractions (percentage of the total volume occupied by each species). To this end we average the packing fraction of each species, i.e., $\eta_i = \rho_i v_i$, along the sedimentation path. Here $v_i$ is the particle volume of species $i$. We compute $\eta_i$ for each set of $\mu_i$ along the path (see Methods) and transform the stacking diagram from the $\bar{\mu}_r - \bar{\mu}_p$ to the $\bar{\eta}_r - \bar{\eta}_p$ plane. In both planes each point of the stacking diagram represents a sedimentation path and therefore directly corresponds to an experimental sample in sedimentation-diffusion-equilibrium.

**Comparison with experiments.** In the experiments[22], stacking sequences with isotropic ($I$), nematic rod-rich ($N_r$), and nematic plate-rich ($N_p$) layers are reported. The samples, reproduced in Fig. 3, were initially prepared with packing fractions $(\bar{\eta}_r, \bar{\eta}_p) = (0.02, 0.18)$ (a), $(0.10, 0.08)$ (b), and $(0.10, 0.01)$ (c).

To compare with the experiments, we find the paths for which the average packing fractions $(\bar{\eta}_r, \bar{\eta}_p)$ and sample height match the experimental values. The height is measured form the pictures knowing that the width of the cuvettes is 10 mm. The buoyant mass ratio, and hence the slope of the path, is the same for all samples. There is an uncertainty of ~25% in the experimental particle dimensions[22,23,44] (note small deviations in the diameter of the rods and the height of the plates greatly affect the particle volumes and therefore the packing fractions). Also, solvent evaporation can occur experimentally, affecting the packing fractions[25]. Hence, to find the paths we fix the composition of the mixture to the experimentally reported value and allow a variation in the total packing fraction. This is equivalent to assuming that an unknown percentage of the solvent has been evaporated (alternatively we could allow a variation in the particle sizes). The best agreement between theory and experiment occurs assuming that 25%(a), 50%(b) and 60%(c) of the solvent evaporated during the long equilibration times. These values are consistent with the position of the meniscus in Fig. 3 (the sample heights are (a) 23 mm, (b) 18 mm, and (c) 17 mm) if the samples were filled to the same height initially. The density and nematic order parameter profiles along the sedimentation paths are shown in Fig. 3. Due to gravity the profiles are inhomogeneous (also within a layer of a given mesophase) in contrast to what happens in a bulk state in absence of gravity. The stacking sequence can be read off directly from the nematic order parameter profiles. In the isotropic layers both order parameter vanish, $S_r = S_p = 0$. In the nematic layers rods and plates orient themselves perpendicular to each other (see schematics in Fig. 1): the dominant species has a positive order parameter $S_i > 0$ (particles aligned along the director) and the minority species has a negative order parameter $S_j < 0$ (particles perpendicular to the director).

Despite the complexity of the experiments and the simplicity of the theory, the agreement of the respective results is excellent. All three stacking sequences, namely (from top to bottom) $IN_p$ Fig. 3a, $IN_rN_p$ Fig. 3b and $IN_r$ Fig. 3c are reproduced. Both the phase identity of each layer and their order in the sequence are correctly predicted. Even the vertical positions of the interfaces between layers agree semi-quantitatively. The density profiles show that the $N_p$ and $N_r$ phases are rich in plates and rods, respectively. Interestingly, the isotropic phase can be either dominated by plates, Fig. 3a, or by rods, Fig. 3b, c. This affects the order that the isotropic layer occupies in the stacking sequence for other values of the buoyant mass ratio as we will see below. The experimental results and the theoretical predictions differ in two aspects. In Fig. 3b the theory overestimates the thickness of the $N_r$ layer. The experimentally reported packing fraction of rods is larger than that of plates and the theory predicts almost perfect demixing between the species, see the packing fraction profiles in Fig. 3b. As a result the predicted $N_r$ layer is thicker than the $N_p$ layer. The opposite, however, is observed in the experiments. The other discrepancy between theory and experiments is the prediction of a thin $N_p$ layer at the bottom of sample 3 which is not observed experimentally, see Fig. 3c. This could be due to interfacial effects that are not considered in the sedimentation path theory approximation. The surface tension associated with the emergence of the new interface might prevent such slim layer to appear in the experiments. Also, for other values of the buoyant mass ratio within the experimental uncertainty, the bottom $N_p$ layer is not present (see Supplementary Fig. 1). Nevertheless, the discrepancies found are certainly not surprising given the simple microscopic theory we use to describe the bulk, the experimental uncertainties in both particle dimensions and masses, as well as other bulk factors not taken into account such as polydispersity.

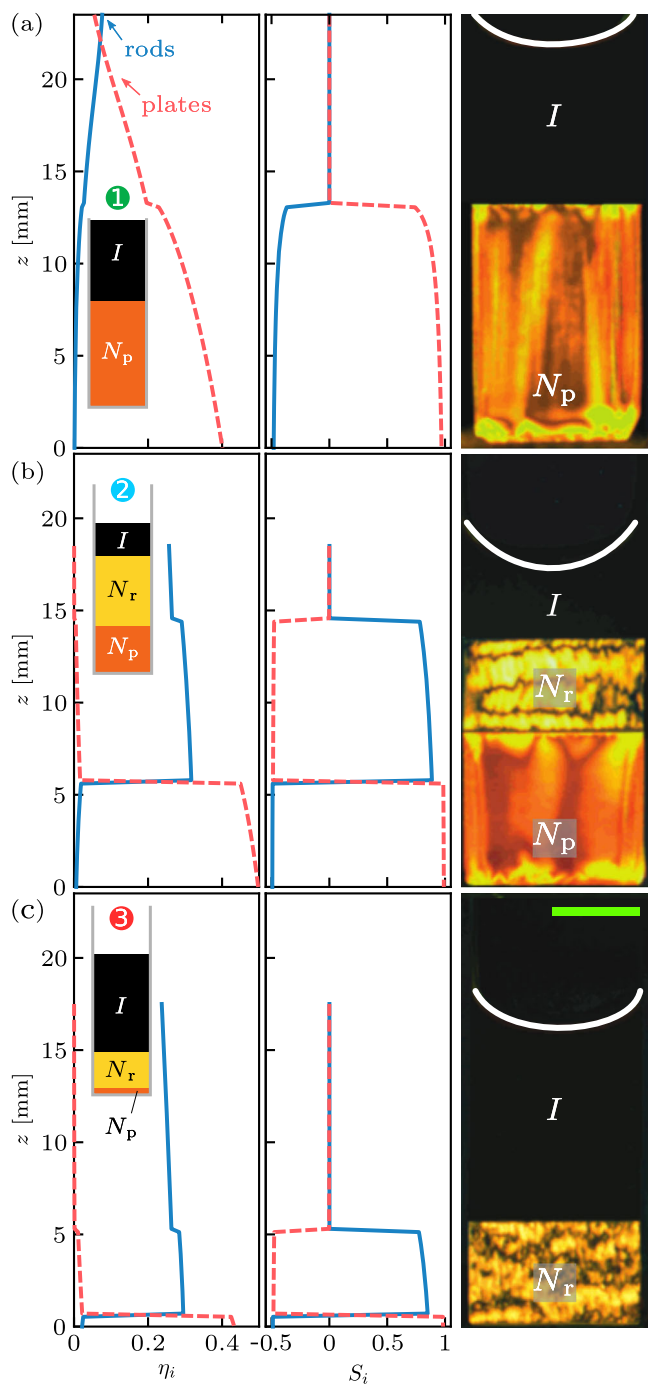

**Fig. 3 Comparison with experiments.** Vertical profiles of the packing fractions $\eta_i$ and the nematic order parameters $S_i$ with $i = r, p$ for rods (solid-blue lines) and plates (dashed-red lines), respectively. Three different samples labeled as 1, 2, and 3 (indicated by the colored circles) are shown. Schematics of the samples showing the isotropic (I), nematic rod-rich ($N_r$) and nematic plate-rich ($N_p$) layers are represented in the insets. The average packing fractions ($\bar{\eta}_r, \bar{\eta}_p$) are (0.027, 0.24) (**a**), (0.20, 0.16) (**b**), and (0.25, 0.025) (**c**). The corresponding experimental samples (pictures taken between crossed polarizers) by van der Kooij and Lekkerkerker[22,23] are also shown (adapted with permission[23], Copyright 2000 American Chemical Society). The meniscus is highlighted with a white line. The bright areas are due to light being polarized in layers with orientational order. The scale bar (green) is 5 mm.

Recall that we adjust only the amount of solvent evaporated to find the best agreement between theory and experiment. Any value of the buoyant mass ratio within the experimental uncertainties reproduces the experimental stacking sequences using only the evaporation as a free parameter in the theory (see Supplementary Fig. 1).

**Bulk and stacking diagrams**. We discuss now the intricate connection between the stacking sequences under gravity and the bulk phase diagram of the mixture. The bulk phase diagram according to our microscopic theory is shown in Fig. 4a, b in the planes of chemical potentials and packing fractions, respectively. To reassure the validity of the theory, we compare the chemical potentials of the $I-N_p$ and $I-N_r$ transitions in monocomponent systems (i.e., $\mu_r \to -\infty$ or $\mu_p \to -\infty$) with those of parallel hard spherocylinders[46,47] and hard cut spheres[48] according to simulations, see the violet arrows in Fig. 4a. We expect that both spherocylinders and cut spheres behave similarly to our cylindrical particles at the relatively low density of the isotropic-nematic transition and large particle anisotropies considered here.

The sedimentation paths of the samples in Fig. 3 are depicted in Fig. 4a. The stacking sequences can be read off by simply following the direction, and observing the binodals or the phase regions crossed by each path. For example, the stacking sequence of sample 1 is $IN_p$ from top to bottom since the path crosses only the $I-N_p$ binodal.

The experimentally observed layers in a stacking sequence do not represent coexisting phases in bulk[21]. For example, the sequence $IN_rN_p$ shown in Fig. 3b should not be interpreted as a triple thermodynamic coexistence between $I$, $N_r$, and $N_p$ bulk phases. In reality the three phases might or might not coexist in bulk, i.e., in the absence of gravity. This is because due to gravity, the sample does not represent a state point in bulk but a set of state points along the sedimentation path. Note also that for typical colloidal particles and sample heights, the paths cover a large region of the bulk phase diagram, see e.g., the paths in Fig. 4a. Therefore, the observation of more than three layers in a stacking sequence does not imply violation of the Gibbs phase rule. Observing a single sample, we can conclude that any two consecutive phases in the sample (e.g., $IN_r$ and $N_rN_p$ in the sequence $IN_rN_p$) coexist in bulk since the path crosses a bulk binodal at the position of the interfaces between two consecutive layers. However, one cannot conclude whether or not two non-consecutive layers (e.g., $I$ and $N_p$ in the sequence $IN_rN_p$) coexist in bulk.

As discussed above, it is useful to group the stacking sequences in a finite height stacking diagram[41]. Figure 4 shows the stacking diagrams for samples with heights $h = 5$ mm (c,d) and 18 mm (e,f) in the plane of average chemical potentials (c,e) and average packing fractions (d,f) along the path. In bulk three binodals meet at a triple point, see Fig. 4a. The stacking diagrams contain sedimentation binodals of type I due to paths that either start or end at a bulk binodal plus one terminal line due to paths crossing the bulk triple point. Six different sequences occur: $I$, $N_p$, $N_r$, $IN_p$, $IN_r$, and $IN_rN_p$. If present in the sequence, the $N_p$ layer is always the bottom layer due to the slope of the path (buoyant mass ratio) being steeper than the slope of each bulk binodal. This behavior is not surprising since the plates are heavier than the rods and hence dominate the bottom phases. The total mass density profile, i.e., the sum of the mass density of the particles and the solvent (see Methods), monotonically increases towards the bottom of the sample in all cases (also if the buoyant masses are negative in which case the particles cream up). This is consistent with Archimedes' principle, although it should be noticed that denser particles can also float on top of a lighter fluid[38,39]. Analog to bulk points at which several phases coexists, such as triple points, there exist points in the stacking diagram at which several stacking sequences meet (due to e.g., the crossing of two sedimentation binodals). A discussion about these points is provided in Supplementary Note 1.

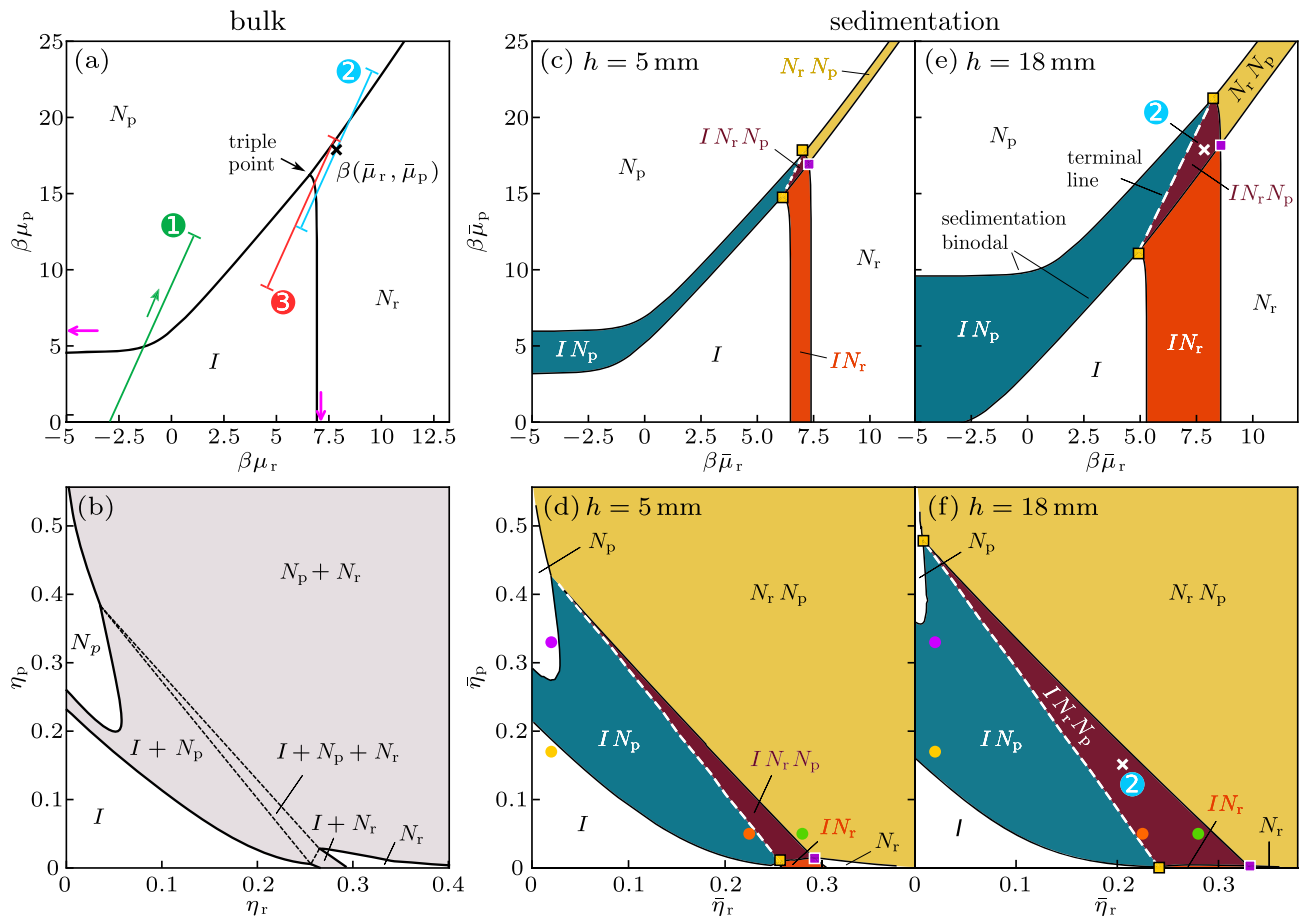

**Fig. 4 Bulk and stacking diagrams.** Bulk phase diagram in the plane of chemical potentials of rods $\mu_r$ and plates $\mu_p$ (**a**) and also in the plane of packing fractions of rods $\eta_r$ and plates $\eta_p$ (**b**). The chemical potentials are scaled with the inverse temperature $\beta = 1/(k_B T)$. Solid lines are the binodals. The stable phases are isotropic (I), nematic plate-rich ($N_p$), and nematic rod-rich ($N_r$). The violet arrows in (**a**) mark the $I - N_p$ and $I - N_r$ transitions of pure systems of plates (horizontal arrow) and rods (vertical) obtained by simulations[46–48]. The line segments are the sedimentation paths of samples 1, 2, and 3. The green arrow illustrates the direction of the paths from top to bottom. The shadow region in (**b**) is the two phase region and the dashed lines are the tie lines of the triple point. Stacking diagrams for heights $h = 5$ mm (**c**, **d**) and 18 mm (**e**, **f**) in the plane of average chemical potentials of the sedimentation paths for rods $\bar{\mu}_r$ and plates $\bar{\mu}_p$ (**c**, **e**) and average packing fractions of rods $\bar{\eta}_r$ and of plates $\bar{\eta}_p$ in the samples (**d**, **f**). Black solid (white dashed) lines are sedimentation binodals (terminal lines). Each stacking sequence is denoted from top to bottom and colored differently (except for the white regions that represent sequences with only one layer). The midpoint of the path for sample 2 is marked with a black cross in (**a**). The position of this sample is also marked in the stacking diagrams with crosses in (**e**) and (**f**). The colored squares indicate the points at which four stacking sequences meet due to the crossing between two sedimentation binodals (violet squares) or the intersection of sedimentation binodals and a terminal line (yellow squares). The colored circles in (**d**) and (**f**) mark samples with the same colloidal packing fractions in both stacking diagrams.

The length of the path in the $\mu_r - \mu_p$ plane is relevant to determine the stacking sequence. For example, varying the length of path 2 in Fig. 4a can alter the stacking sequence from $IN_rN_p$ to $I$, $IN_r$, $N_r$, $N_rN_p$, or $N_p$. Therefore the stacking diagrams are calculated at fixed sample height since the length of the path is proportional to $h$. It is worth noting that for infinitely small sample height the sedimentation path has vanishing length and it is a point in the plane of chemical potentials. Hence, the stacking diagram for $h \to 0$ coincides with the bulk phase diagram. In Fig. 4 it is apparent how the stacking diagram tend to the bulk diagram by decreasing $h$. By increasing $h$ the regions with a single layer sequence shrink in size at the expenses of the regions with multiple layer sequences that are enlarged. This reflects that the longer the path is the more likely it crosses additional binodals. To highlight the importance of the sample height, we indicate by colored circles pairs of illustrative samples with the same colloidal packing fractions but different heights in Fig. 4d and f. In all cases the stacking sequences change upon changing the sample height. For example, for packing fractions $(\bar{\eta}_r, \bar{\eta}_p) = (0.225, 0.05)$ the stacking sequence changes from $IN_p$ if $h = 5$ mm to $IN_rN_p$ if $h = 18$ mm. Therefore, the stacking sequence is not only determined by the colloidal concentration since the occurring layers also depend on the sample height.

For this buoyant mass ratio ($s = m_p/m_r = 3$) the topologies of the stacking and the bulk diagrams are the same in the sense that there is a one-to-one correspondence between bulk regions and stacking sequences for any sample height. For example, the triple point region $I + N_p + N_r$ in Fig. 4b and the region of the stacking sequence $IN_rN_p$ in Fig. 4d, f correspond to each other, although they are different objects. Recall that bulk phases and stacking sequences differ substantially since (i) the order of the layers plays a role in the stacking diagram but not in bulk, and (ii) bulk phases are homogeneous while layers in a stacking diagram are not, see density profiles in Fig. 3. This one-to-one correspondence is not a general feature since the topology of the stacking diagram changes with the buoyant mass ratio.

**Inferring bulk behavior from sedimentation experiments.** Above we have incorporated gravity into a theoretical calculation of bulk phase behavior and compared with experimental samples. Here, we address the experimentally relevant inverse problem. van der Kooij and Lekkerkerker reported also four samples containing layers with liquid-crystalline positional order[23]. These stacking sequences, reproduced in Fig. 5a, contain plate-rich columnar ($C$) and (most likely[23]) rod-rich smectic ($X$) layers. In addition to the orientational order, in the columnar (smectic) mesophase the particles are positionally ordered along two (one) spatial directions. Complex stacking sequences with five distinct layers such as $IN_rXN_pC$ occur. The experimental particles are highly polydisperse which heavily alters the bulk transition densities of phases with positional order[49,50]. Therefore, attempting to extend the density functional to incorporate smectic and columnar phases[51,52] is not a promising route to reproduce the experimental results involving phases with positional order. Instead, we use the experimental sequences to construct the bulk binodals of phases with positional order. Roughly speaking, we disentangle the effects that gravity has on the samples to find the bulk behavior (in absence of gravity).

The slope of the paths remains unchanged and their lengths are obtained by measuring the sample height from the experimental pictures, Fig. 5a. We then construct a bulk phase diagram, see Fig. 5b, with the approximated location of two new binodals, $N_p - C$ and $N_r - X$. We assume the simplest form for the binodals, i.e., horizontal or vertical lines, which is justified since (i) the binodals connect pure transitions in the monocomponent systems to other binodals and (ii) at high density the mixture is expected to be completely segregated. Then, we find the binodal location in the bulk phase diagram together with the position of the sedimentation paths such that both the experimental stacking sequences and the thicknesses of the individual layers are best reproduced. The resulting theoretical stacking sequences are depicted in Fig. 5a for direct comparison with the experiments. All the experimental sequences are reproduced and we can infer the topology of the bulk phase diagram from the given set of sedimentation experiments. Note that any change in the bulk topology (e.g., interchanging the position of the predicted $N_r - N_p - X$ and $N_p - X - C$ triple points) produces a different set of stacking sequences. The thicknesses of the individual layers can be also reproduced quantitatively in most cases. Small deviations occur, especially if the sedimentation path is close to a bulk triple point, due to the simple approximation we use for the binodals. Near triple points the curvatures of the binodals can be large, c.f. the $I - N_p - N_r$ triple point in Fig. 5b, and a straight line is a crude approximation. Nevertheless, with a larger number of samples it might be possible to reproduce the curvature of the bulk binodals and gain further insight into the phase transition.

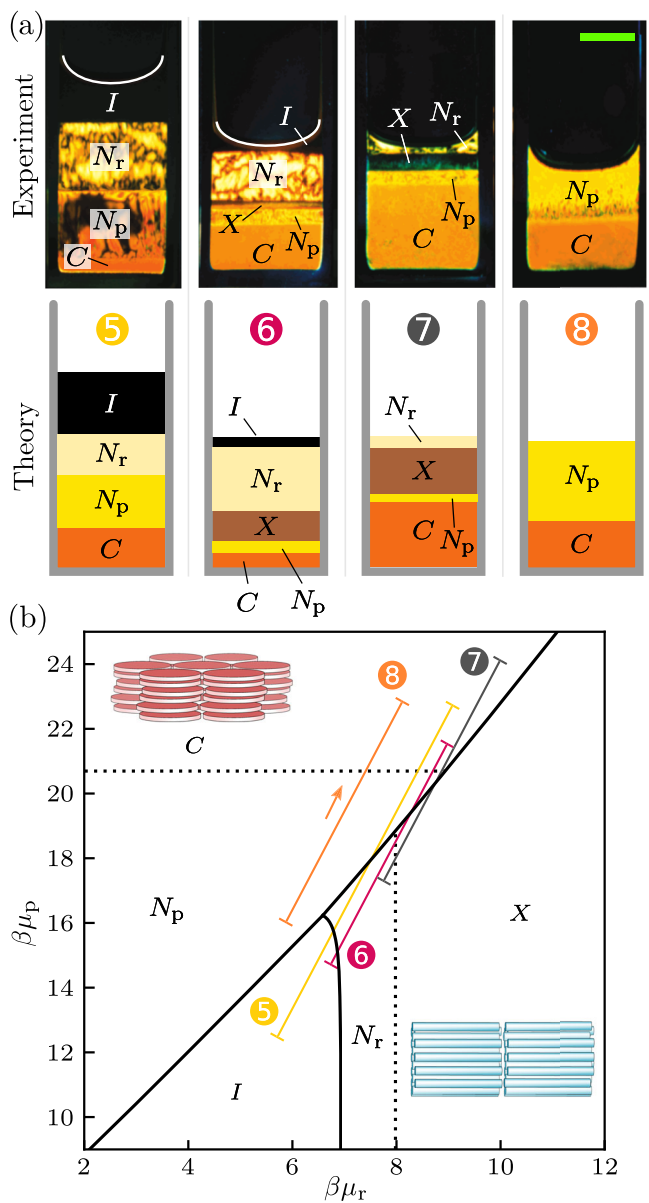

**Fig. 5 Positionally ordered phases. a** Experimental samples and corresponding theoretical predictions of sequences containing isotropic ($I$), nematic plate-rich ($N_p$), and nematic rod-rich ($N_r$) fluid layers as well as layers with positional order: $C$ (columnar plate-rich) and $X$ (smectic rod-rich). The white line indicates the position of the meniscus. The photographs of the experimental samples (adapted with permission[23], Copyright 2000 American Chemical Society) were taken from crossed polarizers. The scale bar (green) is 5 mm. **b** Bulk phase diagram in the plane of chemical potential of rods $\mu_r$ and plates $\mu_p$ for the region next to the triple point. The chemical potentials are scaled with the inverse temperature $\beta = 1/(k_B T)$. Solid lines are the binodals involving isotropic and nematic phases obtained with density functional theory. Dotted lines are the inferred location of the $N_p - C$ (horizontal line) and $N_r - X$ binodals. Schematics of particle arrangement in the $C$ and $X$ phases are shown. The sedimentation paths correspond to the samples shown in (a). The orange arrow in sample 8 indicates the direction of the paths from top to bottom.

Note that the curvature of a binodal in the $\mu_r - \mu_p$ plane is given by the ratio between the density jumps of each species at the phase transition.

Isotropic and columnar phases do not coexist in bulk, Fig. 5b. However, in samples 5 and 6 in Fig. 5a (sequences

$IN_rN_pC$ and $IN_rXN_pC$, respectively) both an isotropic and a (non-adjacent) columnar layer are present. This illustrates that simultaneously occurring layers in a stacking sequence do not need to imply bulk coexistence between the respective phases.

**Changing the buoyant mass ratio.** The topologies of the bulk and the stacking diagrams in Fig. 4b and d are the same (there is a one-to-one correspondence between bulk regions and stacking sequences) for the particular value of the buoyant mass ratio $s = m_p/m_r = 3$. Changing the buoyant mass ratio does not alter the bulk phase diagram but it can modify the topology of the stacking diagram. A change in the buoyant masses can be achieved experimentally by changing the material (inner core[53], coating[54]) of the colloidal particles, and also by changing the solvent density.

To illustrate the effect of changing the buoyant mass ratio, we calculate first the stacking diagram for the idealized case of samples with infinite height[21]. In this limit a sedimentation path is a straight line (not a segment) that can be described with two variables: slope $s$ and intercept $a$ in Eq. (3). A stacking diagram in the $s - a$ plane can be calculated, see Fig. 6a, by locating the paths that form the boundaries between different stacking sequences[21]. These are: paths tangent to bulk binodals, paths that cross triple points, and paths that are parallel to the binodals in the limits $\mu_i \to \pm\infty$. The stacking sequences for finite samples are then given by those in the infinite sample height limit and also by their subsequences formed by removing layers at the top/bottom of the sequence. It becomes apparent from the case of infinite height that the precise value of $s$ is not critical in the sense that it is possible to vary $s$ in a certain range without altering the sequences qualitatively. For example, no qualitative change occurs for buoyant mass ratios $s \gtrsim 2$, see Fig. 6a. This is particularly relevant considering that due to the experimental uncertainties[22], we estimate that the buoyant mass ratio lies within the confidence interval $s \in [2, 5]$ (see Supplementary Fig. 1).

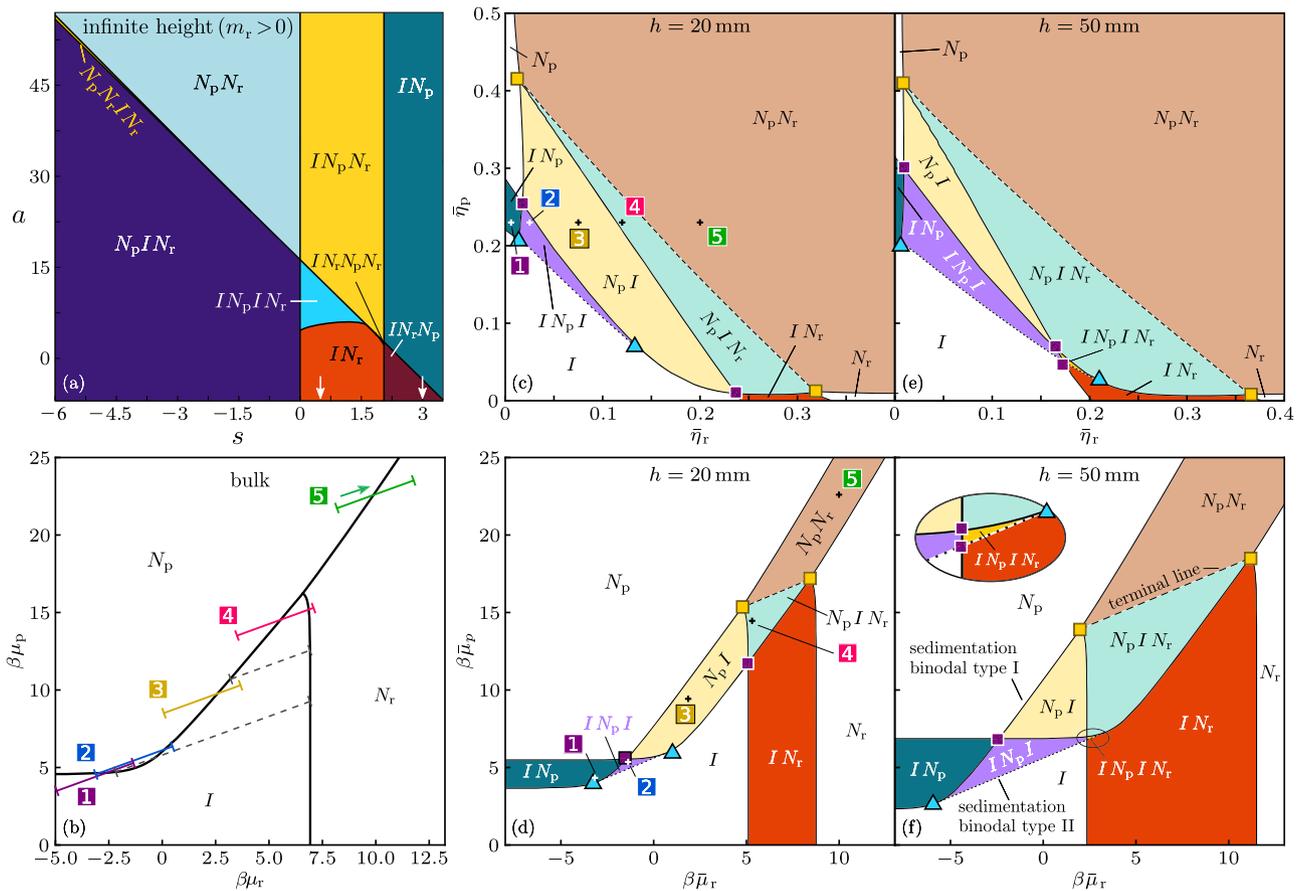

**Fig. 6 Changing the buoyant mass ratio. a** Stacking diagram for samples in the limit of infinite height and positive buoyant mass of rods ($m_r > 0$) in the plane of slope $s = m_p/m_r$ and intercept $a$ of the sedimentation paths. Each region represents a sequence, labeled from top to bottom (the reverse sequences appear if $m_r < 0$). The white arrows highlight the slopes $s = 0.5$ and 3 used here. Bulk phase diagram if the plane of chemical potential of rods $\mu_r$ and plates $\mu_p$ (**b**) with the paths of samples 1 to 5 with $s = 0.5$ and height $h = 20$ mm (solid lines). The chemical potentials are scaled with the inverse temperature $\beta = 1/(k_BT)$. The stable phases are isotropic (I), nematic rod-rich ($N_r$), and nematic plate-rich ($N_p$). The green arrow in sample 5 indicates the direction of all paths from top to bottom. Stacking diagrams for heights $h = 20$ mm (**c, d**) and 50 mm (**d, e**) in the plane of average packing fractions of rods $\bar{\eta}_r$ and plates $\bar{\eta}_p$ (**c, e**) and average chemical potentials of rods $\bar{\mu}_r$ and plates $\bar{\mu}_p$ (**d, f**). Black solid lines are sedimentation binodals of type I (paths that either end or start at a binodal), black-dashed lines are terminal lines (paths crossing the triple point), and black-dotted lines are sedimentation binodals of type II (paths tangent to binodals). Each sequence is denoted from top to bottom and colored differently (except for the white regions that represent sequences with only one layer). The inset in (**f**) is a close view of a small region. The samples 1 to 5 in (**b**) are also shown in (**c**) and (**d**), as indicated. The blue triangles indicate points at which three stacking sequences meet due to the bifurcation of a sedimentation binodal of type II from one of type I. The colored squares in panels (**c, d, e, f**) indicate the points at which four stacking sequences meet due to the crossing between two sedimentation binodals (violet squares) or the intersection of sedimentation binodals and a terminal line (yellow squares). Two sedimentation binodals cross (violet squares) whenever a sedimentation path in bulk simultaneously start and end at a binodal, as illustrated by two gray-dashed paths in (**b**).

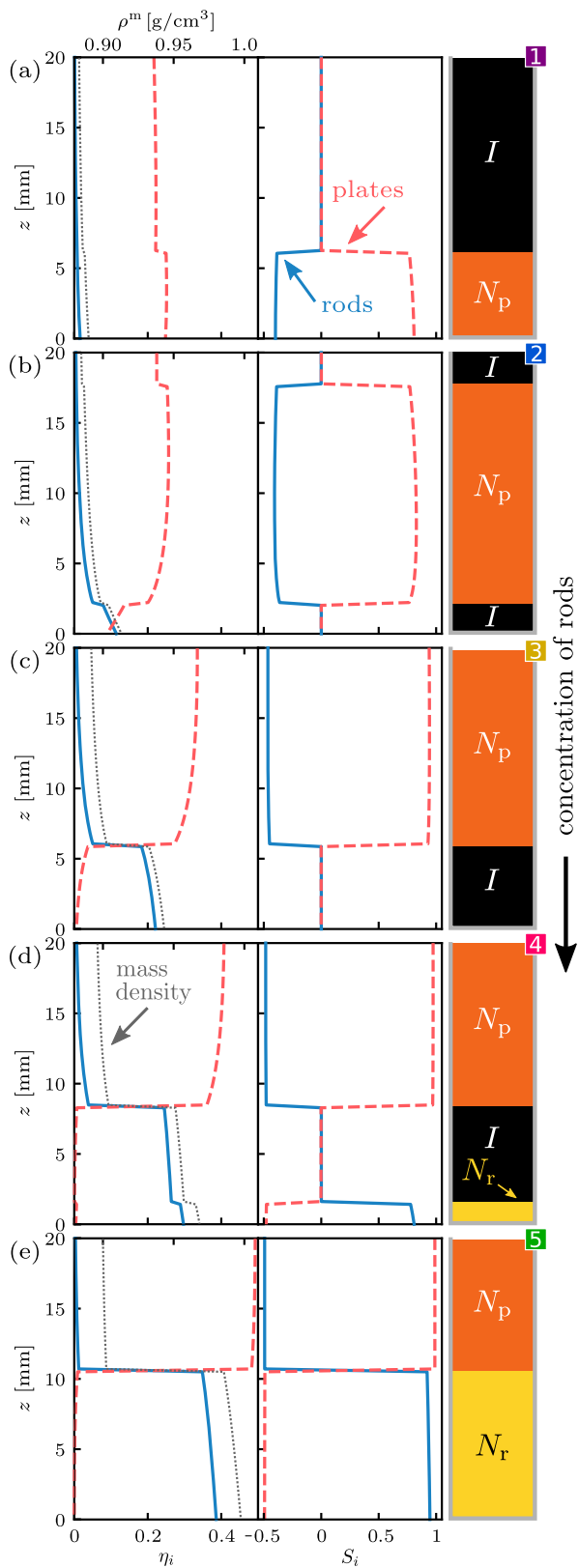

**Fig. 7 Changing the concentration of rods.** Packing fraction $\eta_i$ ($i = $ r, p for rods and plates, respectively) and total mass density $\rho^m$ profiles, uniaxial order parameter $S_i$, $i = $ r, p profiles, and schematics as a function of the vertical coordinate $z$ for samples of height $h = 20$ mm, fixed concentration of plates $\bar{\eta}_p = 0.23$ and varying concentration of rods $\bar{\eta}_r = 0.006$ (**a**), 0.025 (**b**), 0.075 (**c**), 0.12 (**d**), and 0.20 (**e**). The slope of the sedimentation path is $s = 0.5$. The corresponding sedimentation paths are depicted in Fig. 6b. The samples are also marked in the stacking diagrams of Fig. 6c, d using the same labels as here (colored squares with a number from 1 to 5).

A buoyant mass ratio in the interval $s \in [0, 1.75]$ produces richer phenomenology than $s = 3$ (three vs. two distinct stacking sequences). We select a value in that interval, $s = 0.5$, and calculate the experimentally relevant stacking diagrams for finite height. Selected paths with slope $s = 0.5$ are represented in Fig. 6b. The paths correspond to a suspension in toluene of the same rods as before but lighter plates with gravitational length $\xi_p = 11$ mm. The slope is such that there are paths tangent to the $I$–$N_p$ bulk binodal. Those paths create a new boundary in the stacking diagram that increases significantly the number of stacking sequences. We show stacking diagrams in Fig. 6c, d for samples with $h = 20$ mm and in Fig. 6e, f for samples with $h = 50$ mm. Both cases are much richer than the stacking diagrams for $s = 3.0$, cf. Fig. 4. The one-to-one correspondence between bulk regions, Fig. 4b, and stacking sequences, Fig. 6c, e, is lost, emphasizing that bulk and stacking diagrams are fundamentally different objects.

Moreover, the topology of the stacking diagram changes by increasing the sample height due to the occurrence of the complex stacking sequence $IN_pIN_r$ in samples with $h = 50$ mm, Fig. 6e, f. Such a four-layer sequence only occurs for significantly long paths. The path needs to cross the bulk $I$–$N_p$ binodal twice (only possible if $s \in [0, 1.64]$) and the $I$–$N_r$ binodal, see the bulk diagram in Fig. 6b. This gives a lower limit of $h \gtrsim 48$ mm for the occurrence of the four-layer sequence if $s = 0.5$. The topological change of the stacking diagram is driven by a change in the sample height. The complex sequence $IN_pIN_r$ illustrates that the same phase ($I$) can reenter the sequence[25] even though there is no $I$–$I$ demixing in bulk. Instead, the path crosses the $I$–$N_p$ binodal twice.

**Parametric study of stacking sequences and different representations of the stacking diagram.** We investigate five selected samples with the same height, $h = 20$ mm, and packing fraction of plates, $\bar{\eta}_p = 0.23$, but different packing fraction of rods. Their paths are depicted in Fig. 6b and the corresponding state points in the stacking diagram are indicated in Fig. 6c, d. The density and uniaxial profiles are shown in Fig. 7 together with schematics of the stacking sequences.

By increasing $\bar{\eta}_r$ we observe five different stacking sequences $IN_p$, $IN_pI$, $N_pI$, $N_pIN_r$ and $N_pN_r$. These sequences include the formation of bottom isotropic layers, Fig. 7a, b, a floating isotropic layer between two nematic layers, Fig. 7d, and a floating nematic between two isotropic layers, Fig. 7b. The inversion of the sequence $IN_p$ to $N_pI$, see Fig. 7a, c, also occurs by increasing $\bar{\eta}_r$. Such inversion was experimentally observed in a polydisperse suspension of plates[55] by changing the colloidal concentration, and attributed to a pronounced fractionation with respect to plate thickness. The total colloidal packing fraction (or number density) in the isotropic layers is always smaller than it is in the nematic layers, see the density profiles of Fig. 7. However, depending on the bulk region covered by the path, the mass

The value $s = m_p/m_r = 1$ corresponds to the equality of buoyant masses of plates and rods and hence delimits which species is heavier: rods ($s < 1$) or plates ($s > 1$). Interestingly, the stacking diagram for infinite height, Fig. 6a, reveals that which species is the heavier does not play a decisive role since no qualitative change occurs at $s = 1$.

density can be greater in the isotropic layer than in the nematic layer which facilitates the occurrence of bottom isotropic layers. In all cases the total mass density (see Methods) increases monotonically towards the bottom of the sample, see Fig. 7.

A detailed investigation of this complex evolution of sequences can be performed with a representation of the stacking diagram in the plane of packing fraction of rods and vertical coordinate, see Fig. 8a. The diagram indicates the occurrent layer at a given vertical position and concentration of rods. Both the height $h = 20$ mm and the plate concentration $\bar\eta_p = 0.23$ are fixed. Figure 8a shows the introduction of the bottom isotropic layer closely followed by the elimination of the top isotropic layer by increasing $\bar\eta_r$. Then, an $N_r$ layer is introduced which eventually replaces the $I$ layer at the bottom entirely. At high packing fraction of rods, only nematic layers appear and the thickness of the $N_r$ layer increases by increasing $\bar\eta_r$, as expected.

The stacking diagram in the plane of sample height and vertical coordinate at fixed concentrations, $\bar\eta_r \approx 0.18$ and $\bar\eta_p \approx 0.05$, is shown in Fig. 8b. This stacking diagram is relevant for experimental realizations since it represents the creation of several samples that differ only in the sample height. For $h = 47$ mm we observe the sequence $N_p I$ which upon increasing $h$ transforms first into a floating isotropic $N_p I N_r$, followed by the four-layer sequence $I N_p I N_r$. Finally at $h = 53.7$ mm the two isotropic layers merge into a single layer due to the elimination of the $N_p$ layer, which gives rise to the sequence $I N_r$. See schematics of the evolution in Fig. 8c. Such complex behavior involving four different sequences is observed by varying the sample height only by 17%, from 47 mm to 55 mm. Interestingly, in the $h - z$ plane, Fig. 8b, the lower $I - N_p$ boundary is parallel to the top sample-air boundary, whereas the upper $I - N_p$ boundary is horizontal.

It is worth pointing out that varying the sample height at fixed concentrations not only changes the length of the sedimentation path but also its position in the $\mu_r - \mu_p$ plane. This change gives rise to the observed nontrivial dependence of the stacking sequence on the sample height.

## Discussion

To demonstrate the validity of the concept of sedimentation paths[21] we have studied sedimentation-diffusion-equilibrium of a colloidal plate-rod mixture and found excellent quantitative agreement with the well-known experiments conducted by van der Kooij and Lekkerkerker[22,23]. We have shown how to systematically analyze and interpret the stacking sequences observed experimentally, group these sequences in a stacking diagram, and predict the stacking diagram from the bulk phase diagram of the system. Moreover, we have also shown how to infer the bulk phase behavior of the mixture from the experimental results under gravity and have also predicted both a different set of stacking sequences and a complex evolution of the sequences by simply changing the height of the samples. All predictions can be verified experimentally by altering the buoyant masses of the particles and systematically varying the height of the samples.

Some gravity-induced effects, like the formation of a sequence with five layers, were attributed to polydispersity[22] since the occurrence of more than three layers was understood as an apparent violation of the Gibbs phase rule. However, due to gravity it can only be guaranteed that any two consecutive layers that share an interface in the sample coexist in bulk. Hence, as pointed out in other works[21,40], the occurrence of say five layers in a stacking sequence does not imply the existence of a quintuple point in the bulk phase diagram. Such multi-phase bulk coexisting points can exist in binary mixtures[56,57] for specific inter-particle interactions but are unrelated to the occurrence of several layers in sedimentation.

Even though polydispersity is almost unavoidable in experiments, our theory reproduces here the observed stacking sequences semi-quantitatively. Adding polydispersity to sedimentation path theory is, in principle, possible provided that the theoretical description of the bulk also incorporates polydispersity, e.g., via a distribution of particle sizes[49,58].

Our results indicate a nontrivial dependence of the stacking sequences on the sample height. Controlling and varying the sample height is, in principle, simple in experimental realizations and it opens a route to find interesting phenomenology and gain insight into the bulk phase behavior. Analytical ultracentrifugation[59], in which centrifugal forces change the strength of gravity, can be also described with our theory. Changing the strength of gravity is an alternative method to vary the length of the sedimentation path leaving the buoyant mass ratio unaltered.

The topology of the stacking diagrams can change with the buoyant mass ratio. However no qualitative change occurs here around a buoyant mass ratio $s$ of unity (which delimits which species is the heavier). This is likely the case in other asymmetric

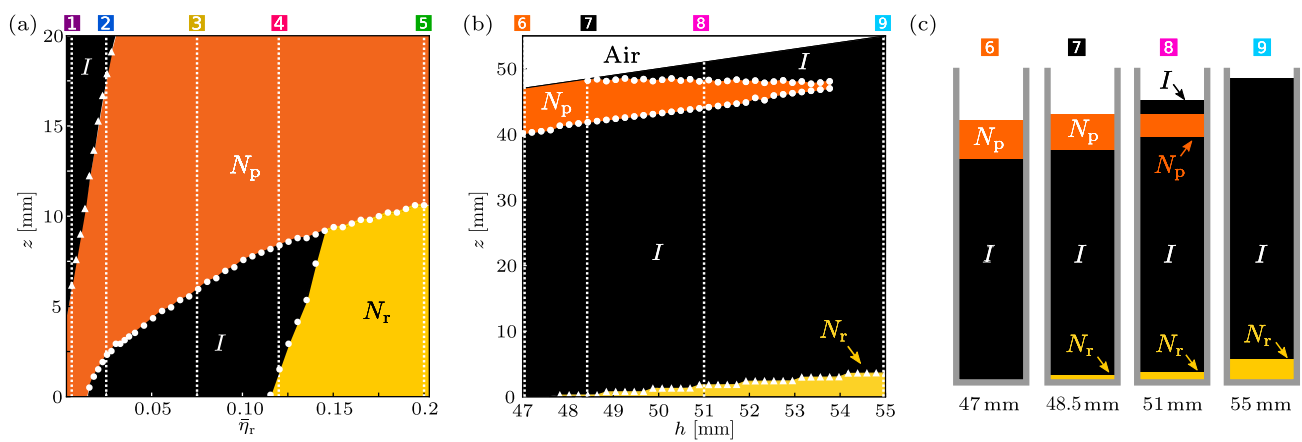

**Fig. 8 Effect of sample height. a** Stable layer at elevation $z$ as a function of the packing fraction of rods $\bar\eta_r$ for samples with fixed height $h = 20$ mm and packing fraction of plates $\bar\eta_p = 0.23$. **b** Stable layer at elevation $z$ as a function of the total sample height $h$ for samples with fixed concentrations $(\bar\eta_r, \bar\eta_p) = (0.177, 0.051)$. The vertical dotted white lines indicate the position in the diagrams of selected samples, labeled by colored squares. **c** Schematics of four selected samples of different heights but identical colloidal concentrations that develop qualitatively different stacking sequences. The layers are isotropic ($I$), nematic plate-rich ($N_p$), and nematic rod-rich ($N_r$).

mixtures. Symmetric mixtures are a special case where $s=1$ can play a special role.

The stacking diagram admits several representations. We have shown two experimentally relevant representations in the plane of vertical interface position between layers, $z$, and either average packing fraction, $\bar{\eta}$, or the overall sample height, $h$. Although these axes are motivated by sedimentation experiments, the $z-h$ plane can be also conceived as a transformation of the bulk phase diagram in the plane of chemical potentials: The vertical position is a linear parameterization of the chemical potential, Eq. (2), and the change in overall height moves nontrivially the path in the plane of chemical potentials implicitly via the constraint of fixed average packing fractions. The inversion of this non-linear integral relationship between average chemical potentials, i.e., the position of the sedimentation path, and the average packing fraction might be another route to obtain the bulk phase diagram from sedimentation experiments.

We expect similar gravity-induced phenomena to occur in other types of mixtures over length scales of a few centimeters provided that the size of at least one species is of the order of 100 nm or larger and there is a significant density difference between the solvent and the particles. For colloidal particles of a few nanometer and micro-emulsions, gravitational effects will be apparent at lengths scales comparable to the gravitational length, which can be of the order of meters. Gravity-induced density gradients also occur in sufficiently large molecular systems and can be also described with sedimentation path theory.

The sedimentation path theory excludes surface effects that occur at the interfaces between two layers and also between the suspension and the bottom/top of the cuvette. Surface effects can be incorporated via a full, spatially resolved, density functional minimization of the inhomogeneous mixture in a gravitational field (see Methods) and might introduce small changes in the position of the boundary lines in the stacking diagram.

The sedimentation path theory and the methodology presented here are general and constitute the basis for carrying out and interpreting sedimentation experiments on both colloid-colloid and polymer-colloid mixtures[60]. Carrying out new experiments to investigate the role of the sample height and that of the buoyant mass ratio would be particularly valuable and enlightening.

## Methods

**Bulk phase behavior.** We use classical density functional theory (DFT)[61] to obtain the bulk phase diagram of the plate-rod mixture. The total free energy $F$ is the sum of the ideal and the excess part ($F = F^{\text{id}} + F^{\text{exc}}$). The ideal contribution to the intrinsic Helmholtz free energy at temperature $T$ for a mixture is given exactly by

$$\beta F^{\text{id}} = \sum_i \int d\mathbf{r} \int d\boldsymbol{\omega} \rho_i(\mathbf{r},\boldsymbol{\omega})\left[\ln\left(\rho_i(\mathbf{r},\boldsymbol{\omega})\Lambda_i^3\right) - 1\right], \quad (4)$$

where $\beta = 1/(k_B T)$ with Boltzmann's constant $k_B$, the sum runs over both species, $\Lambda_i$ is the thermal wavelength of species $i = r, p$, and $\rho_i(\mathbf{r},\boldsymbol{\omega})$ is the one-body density profile of species $i$ at position $\mathbf{r}$ and orientation specified by the unit vector $\boldsymbol{\omega}$. Since we consider only phases without positional order we average out all positions $\mathbf{r}$ and introduce the angular distribution function $\psi_i$ of species $i$ via $\rho_i(\mathbf{r},\boldsymbol{\omega}) = \rho_i \psi_i(\boldsymbol{\omega})$ and normalization $\int d\boldsymbol{\omega}\psi_i(\boldsymbol{\omega}) = 1$. Hence, $\rho_i$ is the number density of species $i$ and we obtain

$$\frac{\beta F^{\text{id}}}{N} = \sum_i x_i \int d\boldsymbol{\omega}\psi_i(\boldsymbol{\omega})\left[\ln\left(\psi_i(\boldsymbol{\omega})\rho_i\Lambda_i^3\right) - 1\right], \quad (5)$$

where $N$ is the total number of particles in the system and $x_i$ is the composition of species $i$.

We use an extended Onsager approximation for the excess (over ideal) contribution to the free energy

$$\frac{\beta F^{\text{exc}}}{N} = \Psi(\eta)\rho \sum_{i,j} x_i x_j \int d\boldsymbol{\omega} \int d\boldsymbol{\omega}' \psi_i(\boldsymbol{\omega})\psi_j(\boldsymbol{\omega}')V_{i,j}^{\text{ex}}(\boldsymbol{\omega},\boldsymbol{\omega}'), \quad (6)$$

with total density $\rho = \sum_i \rho_i$ and $V_{i,j}^{\text{ex}}(\boldsymbol{\omega},\boldsymbol{\omega}')$ being the excluded volume (i.e., the volume inaccessible to one particle due to the presence of another particle) between particles of species $i$ and $j$ with orientations $\boldsymbol{\omega}$ and $\boldsymbol{\omega}'$, respectively. Both rods and plates are modeled as hard cylinders (see Fig. 1) for which analytical expressions for the excluded volume exists[62,63]. To speed up the computation, the azimuthal angle $\varphi$ of both species is averaged over in advance and only the polar dependence, $V_{i,j}^{\text{ex}}(\theta, \theta')$, is retained. Accordingly, we consider only the polar dependence of the angular distribution function $\int_0^{2\pi} d\varphi \psi_i(\boldsymbol{\omega}) = 2\pi \psi_i(\theta)$. This prevents the study of biaxial phases that, on the other hand, are not stable for the particle aspect ratios considered here[64] and are also not observed in the experiments[22,23].

Following Parsons[65] and Lee[66], we replace the prefactor 1/2 in front of the second virial coefficient in Onsager's original expression[62] by

$$\Psi(\eta) = \frac{4 - 3\eta}{8(1 - \eta)^2}, \quad (7)$$

in Eq. (6), which corresponds to the excess free energy per particle of a system of hard spheres according to the Carnahan-Starling equation of state[67]. Here $\eta = \rho\sum_i x_i v_i = \sum_i \eta_i$ is the total packing fraction across all species with $v_i$ being the particle volume of species $i$. The scaling, Eq. (7), does not alter the topology of the phase diagram and serves to improve substantially the agreement of the transition densities compared to computer simulations[64]. In the low density limit ($\eta \to 0$) the original Onsager expression, based on the second virial coefficient, is recovered since $\Psi(\eta \to 0) = 1/2$.

**Minimization.** We perform a free minimization of the functional discretizing $\psi_i(\theta)$ on a one dimensional grid with 160 bins and calculate the uniaxial order parameter of species $i$ according to

$$S_i = \int d\theta \frac{3\cos^2(\theta) - 1}{2} \psi_i(\theta), \quad (8)$$

where the angle $\theta$ is measured with respect to the director of one of the species.

**Bulk coexistence.** The bulk phase diagram is obtained via numerical minimization of the Gibbs free energy per particle

$$g_b = \frac{F}{N} + \frac{P}{\rho}, \quad (9)$$

where $P$ is the osmotic pressure and the total number density is $\rho = \rho_r + \rho_p$ with $\rho_i$ the number density of species $i = r, p$. Mechanical and thermal phase equilibria are fulfilled in the Gibbs ensemble by construction ($P$ and $T$ are fixed). To find chemical equilibrium and hence phase coexistence we search for a common-tangent construction on $g_b(x_r)$, with $x_i = \rho_i/\rho$ the composition of species $i$. The common tangent is equivalent to the equality of chemical potentials of both species in the coexisting phases since $g_b = \mu_r x_r + \mu_p x_p$, with $\mu_i$ the chemical potential species $i$. Hence, for fixed values of $P$, $T$, and $x_r$ (which also fixes $x_p$ since $x_p = 1 - x_r$), we numerically minimize the Gibbs free energy per particle with respect to the total density and the orientational distribution functions of both species and then search for a common tangent.

**Average packing fractions along a path.** To obtain $\eta_i$ along a sedimentation path we minimize the grand canonical potential $\Omega$ per unit of volume

$$\frac{\Omega}{V} = \frac{F}{V} - \rho \sum_i \mu_i x_i, \quad (10)$$

at fixed chemical potentials, $\mu_i$, with respect to the variables $\psi_i$, $x_i$ and $\rho$.

Given the coordinates $\bar{\mu}_i$ of a path, it is straight forward to obtain the corresponding $\bar{\eta}_i$. The opposite procedure, which we use to find the paths that correspond to the experimental samples in Fig. 3, is however involved. Provided with the average packing fractions of a given sample, $\bar{\eta}_i$ we numerically solve a set of non-linear equations to find the corresponding sedimentation path in the plane of chemical potentials.

**Full minimization of the grand potential.** As an alternative to the sedimentation path theory, it would be possible find the solution to the full inhomogeneous system by minimizing the grand potential functional

$$\Omega[\{\rho_i\}] = F[\{\rho_i\}] - \sum_i \int d\mathbf{r} \int d\boldsymbol{\omega} \rho_i(\mathbf{r},\boldsymbol{\omega})(V_i^{\text{ext}}(\mathbf{r},\boldsymbol{\omega}) - \mu_i), \quad (11)$$

with respect to the density profiles of both species $\rho_p(\mathbf{r},\boldsymbol{\omega})$ and $\rho_r(\mathbf{r},\boldsymbol{\omega})$. Here, $F[\{\rho_i\}]$ is the intrinsic free energy functional of the inhomogeneous system. For a gravitational field, the external field is simply $V_i^{\text{ext}} = m_i g z$.

Within the sedimentation path theory, instead of minimizing Eq. (11), we minimize at each value of $z$ along the sedimentation path a corresponding bulk system, Eq. (10), with a height dependent chemical potential given by Eq. (2). The linear dependence of the local chemical potential with the vertical coordinate, Eq. (2), is a direct consequence of the linear dependency of the external potential on $z$. Note, however, that the dependence of the local chemical potentials on other variables such as the composition is more complex since it ultimately depends on the interparticle interactions via the free energy $F$.

**Total mass density**. The total mass density profile, $\rho^m(z)$ is the sum of the mass density of each species plus the density of the solvent. Using the buoyant masses $m_i$, the total mass density is simply[68]

$$\rho^m(z) = \sum_i m_i \rho_i(z) + \rho_s, \quad (12)$$

where $\rho_i(z)$ is the number density of species $i$ at position $z$ and $\rho_s$ is the mass density of the solvent.

## Data availability

All the data supporting the findings are available from the corresponding author upon reasonable request.

Received: 20 April 2021; Accepted: 10 August 2021;
Published online: 07 September 2021

### Acknowledgements
This work is supported by the German Research Foundation (DFG) via project number 436306241.

### Author contributions
T.E. carried out the calculations. T.E., M.S., and D.d.l.H. conceived and designed the concept, and wrote the manuscript.

### Funding
Open Access funding enabled and organized by Projekt DEAL.

### Competing interests
The authors declare no competing interests.

### Additional information
**Supplementary information** The online version contains supplementary material available at https://doi.org/10.1038/s42005-021-00706-0.

**Correspondence** and requests for materials should be addressed to M.S. or D.d.l H.

**Peer review information** *Communications Physics* thanks the anonymous reviewers for their contribution to the peer review of this work. Peer reviewer reports are available.

**Reprints and permission information** is available at http://www.nature.com/reprints

**Publisher's note** Springer Nature remains neutral with regard to jurisdictional claims in published maps and institutional affiliations.

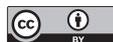